\documentclass{JHEP3}

\usepackage{amsmath}
\usepackage{graphicx}

\newcommand{\mcc}{\mathcal{C}}
\newcommand{\mcl}{\mathcal{L}}
\newcommand{\mco}{\mathcal{O}}
\newcommand{\mcs}{\mathcal{S}}
\DeclareMathOperator{\diag}{diag}

\newcommand{\Ead}{\frac{\dot a}{a}}
\newcommand{\Eadd}{\frac{\ddot a}{a}}
\newcommand{\Eap}{\frac{a'}{a}}
\newcommand{\Eapp}{\frac{a''}{a}}
\newcommand{\Ebd}{\frac{\dot b}{b}}
\newcommand{\Ebdd}{\frac{\ddot b}{b}}
\newcommand{\Ebp}{\frac{b'}{b}}
\newcommand{\End}{\frac{\dot n}{n}}
\newcommand{\Enp}{\frac{n'}{n}}
\newcommand{\Enpp}{\frac{n''}{n}}

\title{Extra-dimensional cosmology with domain-wall branes}
\author{Damien P. George\\%
School of Physics, The University of Melbourne, Victoria 3010, Australia\\%
\email{d.george@pgrad.unimelb.edu.au}}
\author{Mark Trodden\\%
Department of Physics, Syracuse University, Syracuse, NY 13244, USA.\\%
Department of Astronomy, Cornell University, Ithaca, NY 14853, USA.\\%
Department of Physics and Astronomy, University of Pennsylvania, Philadelphia, PA 19104, USA.\\%
\email{trodden@physics.syr.edu}}
\author{Raymond R. Volkas\\%
School of Physics, The University of Melbourne, Victoria 3010, Australia\\%
\email{raymondv@unimelb.edu.au}}
\date{\today}

\abstract{
We show how to define a consistent braneworld cosmology in a model in
which the brane is constructed as a field-theoretic domain wall of
finite thickness.  The Friedmann, Robertson-Walker metric is recovered in
the region of the brane, but, remarkably, with scale factor that depends
on particle energy and on particle species, constituting a breakdown of
the weak equivalence principle on sufficiently small scales.  This
unusual effect comes from the extended nature of particles confined to
a domain-wall brane, and the fact that they feel an ``average'' of the
bulk spacetime.  We demonstrate how to recover the standard results of
brane cosmology in the infinitely-thin brane limit, and comment on
how our results have the potential to place bounds on parameters such as
the thickness of domain-wall braneworlds.
}


\begin{document}


\section{Introduction}

Extra-dimensional theories of particle physics beyond the standard
model have, in recent years, become a standard part of the array of
phenomenological models that we hope to test with the imminent
operation of the Large Hadron Collider (LHC). Some models are extended versions
of the original Kaluza-Klein (KK) construction, in which the known matter
fields propagate in a higher dimensional space, and our usual low-energy
physics is augmented by the properties and interactions of their KK
towers. More dramatically, inspired by the D-branes of string theory,
Arkani-Hamed, Dimopoulos and Dvali~\cite{ADD} (see also~\cite{Antoniadis1990,AADD})
and Randall and Sundrum (RS)~\cite{RS1,RS2} have suggested an alternative
set of extra dimensional constructions, in
which a restricted subclass of fields (often just gravity) propagate in
the additional directions, while standard model particles remain confined
to a $3+1$ submanifold, or {\it brane}
(see also~\cite{Gibbons:1986wg,Visser:1985qm}).

In such models, the brane is typically modelled as infinitely thin -- a
delta-distribution source -- and one then constructs a four-dimensional
effective field theory of the standard model particles plus those propagating
in the bulk, with their associated KK partners. A rich phenomenology
results, with details depending on the particular type of extra dimensional
construction employed (see~\cite{Kaloper:2000jb,Starkman:2000dy,Starkman:2001xu,Nasri:2002rx} for
alternate constructions, and~\cite{Rubakov:2001kp,Langlois:2002bb} for general
overviews.)  Predictions include a spectrum of KK particles expected to be
produced at colliders and modifications of the Newtonian inverse square law
at sufficiently small distances.  Details regarding cosmological evolution
were worked out in the initial series of
papers~\cite{Binetruy:1999ut,Csaki1999,Cline1999,Chung:1999zs,Binetruy:1999hy,Flanagan:1999cu}, where it
was shown that there are deviations from the standard cosmology at
temperatures above a particular {\it normalcy temperature}.

Over the decade following these discoveries, an immense body of work has
been performed to tease out the particle physics and cosmological signatures
of these models.  Cosmology in particular has received much
attention~\cite{Brax:2004xh}, with a great deal of the work focused on inflation,
gravity perturbations and similar cosmological consequences of braneworlds
which would be evident today.  Attention has also been given to more exotic
phenomena, such as colliding
branes~\cite{Takamizu:2004rq,Gibbons:2006ge,Takamizu:2007ks,Saffin:2007qa,Saffin:2007ja},
which is a unique aspect of brane cosmology that can have drastic effects
on the very early universe.
Upcoming precision experiments and missions in
these fields, along with particle physics results from the LHC, promise
increasingly stringent tests in the near future. Such precision allows us to probe
ever more fine detail of extra dimensional theories, requiring an examination
of the sensitivity of the tests to the specifics of the underlying construction.
One such detail is the delta-distribution nature of the brane itself.

It is certainly possible that the $3$-brane on which we find ourselves is an
actual D-brane (or orientifold plane), and that our universe is embedded in
a string-theoretic realisation of, for example, the Randall-Sundrum braneworld.
In that case, it would seem sensible to treat the brane as infinitely thin for
all purposes of low-energy particle physics and cosmology.
If it is necessary to go beyond this approximation and model the effects of a
finite-thickness fundamental brane, then one can do so by averaging the relevant
cosmological quantities over the extent of the brane, and imposing boundary
matching conditions where the thick brane meets the
bulk~\cite{Cvetic:2008gu,Kanti:1999sz,Mounaix2002,Navarro:2005uq}.

However, there is
another possibility, that the braneworld model itself is unrelated to string
theory, but is instead a field-theoretic construction, in which the brane is a
topological defect (a domain wall, for example) in a higher dimensional
space~\cite{Rubakov:1983bb,Akama:1982jy}.
Generally, a scalar field is responsible for generating the topological defect,
and the brane and Randall-Sundrum warped metric become smoothed out versions of
their counterparts in the fundamental
case~\cite{DeWolfe:1999cp,Gremm:1999pj,Csaki:2000fc,Kehagias:2000au,Bazeia:2007nd}.
In such scenarios, the extended nature of the brane in the extra dimensions may have
important implications for phenomenology and cosmology, both through its
complicated relationship to the bulk, and through the field theory mechanisms
through which it must confine the standard model particles.

In a series of papers, two of us (DPG and RRV) have examined this question in
detail for the case in which the brane is a codimension one soliton (a domain
wall) in $4+1$ spacetime dimensions, providing a Minkowski background for
standard model fields. In such a construction~\cite{Davies:2007xr} (see
also~\cite{Davidson:2007cf} for an extension to a larger grand-unified group), a
subtle interplay between the bulk gauge fields (and gauge group), the bulk scalar
fields, and the symmetry breaking field that forms the domain wall is necessary
for standard model fermions, the Higgs field, and the $\mathop{\rm SU}(3)\times \mathop{\rm SU}(2)\times \mathop{\rm {}U}(1)$
gauge bosons to confine and appear in the resulting $3+1$ dimensional effective theory.

In this paper, our goal is to extend previous work to understand the cosmology
of field-theoretic braneworld models, requiring us to abandon the assumption
of Minkowski spacetime on the brane, and to develop a framework in which the
standard model fields feel a background Friedmann, Robertson-Walker (FRW)
metric.  Related work on thick fundamental brane
cosmology~\cite{Kanti:1999sz,Mounaix2002,Navarro:2005uq},
and the prescription for averaging
5d quantities to produce the corresponding effective 4d quantities, is
drastically modified in the case of a domain-wall brane.  As we shall see,
the KK expansion of 5d fields and the integration over the full extent of
the extra dimension leads to some interesting and unexpected features.

The structure of this paper is as follows. In the next section we review
the main points of brane cosmology as constructed around a delta-distribution
source. In Sec.~\ref{sec:domainwallbrane} we then demonstrate how to
build the brane from a field-theoretic domain wall, and discuss how to
define the metric on such an object. We then show how scalar fields and
fermions are confined to such a domain wall, and derive expressions for
the cosmological metrics they experience. The thin wall limit is verified
in Sec.~\ref{sec:thinwall} and we then collect and discuss our main results.


\section{Fundamental brane cosmology}
\label{sec:fund}

In this section we summarise previous results for the cosmological
evolution of a fundamental brane with localised sources
(see~\cite{Binetruy:1999ut,Csaki1999,Binetruy:1999hy} for details,
and~\cite{Chatillon:2006vw} for an extension to higher codimension.)
The idea is to take a 5-dimensional bulk spacetime, include brane, bulk and
brane-localised stress-energy sources, and solve the 5-dimensional Einstein
equations.  The brane is considered to be a fundamental object and is
modelled by a delta distribution, with the total action being  
\begin{equation}
\mcs=\int d^5x \sqrt{-g} \ M_5^3 (R - 2\Lambda)
    \ + \int d^5x \sqrt{-g_\text{brane}} \ \delta(y) \mcl_\text{brane} \ ,
\end{equation}
where $M_5$ is the 5-dimensional gravitational mass scale, $\Lambda$
is the bulk cosmological constant and $g$ and $g_\text{brane}$ are the
determinants of the metric in the bulk and on the brane respectively.
The delta-function localises $\mcl_\text{brane}$ to the brane, which
includes the brane tension and standard model fields.

Since we are interested in cosmological solutions, we consider sources that are homogeneous
and isotropic in the three spatial dimensions, and the most general metric consistent with these symmetries
\begin{equation}
\label{eq:cosmo-metric}
ds_5^2 = -n^2(t,y) dt^2 + a^2(t,y) \gamma_{ij} dx^i dx^j + b^2(t,y) dy^2 \ ,
\end{equation}
where $i,j$ run over the three spatial dimensions, and
$\gamma_{ij}$ is the metric of the three-space, which may be
positively curved, flat or negatively curved.
The Einstein equation is
\begin{equation}
R_{MN}-\frac{1}{2}g_{MN}R + g_{MN} \Lambda = \frac{1}{2M_5^3}T_{MN}  \ ,
\end{equation}
where $M,N$ are 5-dimensional indices.  For explicit expressions of
the components of the Einstein tensor, see Appendix~\ref{app:einstein}.
The brane tension and brane-localised sources are represented jointly
by the 4-dimensional density $\rho_b$ and pressure $p_b$ and appear
in the stress-energy tensor as
\begin{equation}
T^M_{\phantom{M}N} = \delta(by) \diag(-\rho_b, p_b, p_b, p_b, 0) \ .
\end{equation}

For general values of $\Lambda$, $\rho_b$ and $p_b$, Einstein's
equations will yield time-dependent solutions, corresponding, for
example, to an expanding spacetime on the brane.  Before
exploring such solutions, we note that it is possible to fine tune
the sources to produce a time-independent background.  This
corresponds exactly to the scenario of Randall and
Sundrum~\cite{RS1,RS2} who demonstrated that gravity is
localised to a fundamental brane tuned in such a way.

The specific choice necessary is that the brane source be pure tension (corresponding to a 4d cosmological
constant) $\rho_b = -p_b = \sigma$, tuned against the bulk cosmological constant according to
\begin{equation}
\label{eq:fine-tune}
\sigma = \sqrt{-24M_5^6\Lambda} \ .
\end{equation}
Note that this implies that the bulk geometry must be five-dimensional Anti-de
Sitter space AdS$_5$, with $\Lambda<0$.  The corresponding metric solution is then
\begin{equation}
\label{eq:rs-metric}
ds_5^2 = e^{-2\mu |y|}(-dt^2 + \gamma_{ij} dx^i dx^j) + dy^2 \ ,
\end{equation}
where
\begin{equation}
\label{eq:mu-defn}
\mu \equiv \sqrt{\frac{-\Lambda}{6}} = \frac{\sigma}{12M_5^3} \ .
\end{equation}

Moving back to the general time-dependent case, one solves the 5-dimensional
Einstein equations by imposing the Israel matching conditions -- calculating the
discontinuities in the derivatives of the
metric components across $y=0$, and relating these to the delta-distribution
sources (see~\cite{Binetruy:1999ut}).  It turns out that the
behaviours of $\rho_b$, $p_b$ and the metric components evaluated
at $y=0$ are independent of the metric solutions in the bulk, and obey
\begin{align}
\label{eq:eff-conserve}
& \dot\rho_b + 3 H_0(\rho_b + p_b) = 0, \\
\label{eq:eff-friedmann}
& H_0^2 = \frac{\rho_b^2}{144M_5^6} + \frac{\Lambda}{6} - \frac{k}{a_0^2} + \frac{\mcc}{a_0^4} \ ,
\end{align}
where the parameter $k$ takes values $+1,0,-1$ according to whether the metric $\gamma_{ij}$ describes
positively-curved, flat, or negatively-curved spatial 3-sections.
Here a dot corresponds to a time derivative and time has been rescaled
such that $n_0(t)\equiv n(t,y=0)=1$.  The integration constant
$\mcc$ represents an effective radiation term, or so called ``dark
radiation'' (see \cite{Binetruy:1999hy} for bounds on this term from
nucleosynthesis) and from now on we will set $\mcc=0$.  

The effective
4-dimensional scale factor $a_0$ is the 5-dimensional metric component $a(t,y)$ evaluated 
on the brane $a_0(t)\equiv a(t,y=0)$ and $H_0$ is the corresponding Hubble parameter.   
Equation~\eqref{eq:eff-conserve} describes the usual 4-dimensional
conservation of energy on the brane (the continuity equation) and
Eq.~\eqref{eq:eff-friedmann} is the modified Friedmann
equation.  Due to the proportionality $H_0 \sim \rho_b$ instead
of the usual $H_0 \sim \sqrt{\rho_b}$, this Friedmann equation seems
at odds with observation.  The clue to fixing this problem comes
from considering the time-independent Randall-Sundrum solution,
where the brane tension contributed an energy that exactly
cancelled the bulk cosmological constant.  Guided by this, one
writes the total brane source $\rho_b$, $p_b$ as a sum of a
background brane tension $\sigma$ and some other general brane
source $\rho$, $p$
\begin{align}
\rho_b &= \sigma + \rho \ , \\
p_b &= -\sigma + p \ ,
\end{align}
where $\sigma$ is defined as in~\eqref{eq:fine-tune}.  The
effective Friedmann equation for $a_0$ now reads
\begin{equation}
\label{eq:eff-friedmann-2}
H_0^2 = \frac{\sigma}{72M_5^6} \rho + \frac{1}{144M_5^6} \rho^2 - \frac{k}{a_0^2} \ .
\end{equation}
If we assume $\rho$ is small compared to $\sigma$, then the
$\rho^2$ term gives small corrections to the usual behaviour and the
evolution of $a_0$ is driven to first order by $\rho$, with the
proportionality constant playing the role of the effective Planck mass via
\begin{equation}
\label{eq:mp}
M_P^2 \equiv \frac{12M_5^6}{\sigma} = \frac{6M_5^3}{\sqrt{-6\Lambda}} \ .
\end{equation}

An important feature of cosmology in codimension-1 braneworlds is that this entire analysis is independent of the 
behaviour of the metric components in the bulk.  Nevertheless, it is possible to find bulk solutions, and since we will make use of them in a later section we provide them here.  For $\mathcal{C}=0$ they
read~\cite{Binetruy:1999hy,Langlois:2002bb}
\begin{equation}
\label{eq:metric-bulk}
\begin{aligned}
n(t,y) &= e^{-\mu |y|} - \tilde\epsilon\sinh (\mu |y|) \ , \\
a(t,y) &= a_0(t) [e^{-\mu |y|} - \epsilon\sinh (\mu |y|)] \ , \\
b(t,y) &= 1 \ ,
\end{aligned}
\end{equation}
with $\mu$ defined as in~\eqref{eq:mu-defn} and
\begin{align}
\epsilon &\equiv \frac{\rho}{\sigma} \ , \\
\tilde\epsilon &\equiv \epsilon + \frac{\dot\epsilon}{H_0} \ .
\end{align}
Note that for $\epsilon=0$ and $k=0$ we recover the RS
warped-metric solution given by Eq.~\eqref{eq:rs-metric}.

The parameter $\epsilon$ measures the energy density in
matter and radiation, relative to the tension of the brane.  In
terms of this parameter, the Friedmann
equation~\eqref{eq:eff-friedmann-2} is
\begin{equation}
\label{eq:eff-friedmann-3}
H_0^2 = \frac{1}{6M_P^2} \left( 1 + \frac{\epsilon}{2} \right) \rho - \frac{k}{a_0^2} \ ,
\end{equation}
demonstrating that $\epsilon \ll 1$ is required to recover conventional
cosmology. The earliest direct evidence for our standard cosmological
evolution is provided by primordial (big bang) nucleosynthesis (BBN),
which takes place at temperatures of order an MeV. Therefore, we are safe
from cosmological constraints if we choose $\sigma \gg (1 {\rm MeV})^4$.

Finally, we briefly discuss the extension of these results to fundamental
branes with finite thickness~\cite{Kanti:1999sz,Mounaix2002,Navarro:2005uq}.
In these
scenarios the brane and brane-localised sources are modelled as stress-energies
distributed over the finite thickness of
the brane.  The effective 4d quantities, such as the scale factor,
energy density and pressure, are defined to be the spatial average,
over the extent of the brane in the extra dimension, of their
corresponding 5d quantities.  One then rewrites Einstein's equations
in terms of these averaged quantities and identifies corrections to the
infinitely-thin brane scenario.  This averaging prescription is an
important first step in understanding cosmology away from the
infinitely-thin brane limit. However, a more complete treatment is 
essential, since, for
example, in the Minkowski domain-wall set-up, one needs to expand the 5d
fields in KK modes and integrate over the full extent of the extra
dimension.  The rest of this paper is devoted to the development of a
more complete averaging framework, within which it is possible to
analyse the cosmology of domain-wall brane scenarios.


\section{The extension to a domain-wall brane}
\label{sec:domainwallbrane}

Our main goal in this paper is to extend the analysis of the
previous section to the case in which the brane is topological defect -- a domain wall
generated by a scalar field.  The central problem is how to identify the effective 4-dimensional 
scale factor (the analogue of $a_0$) and the equations that describe its time evolution.  As we shall
see, this question turns out to have an interesting and nontrivial resolution, which may have specific
implications for the signatures of such field-theoretic braneworlds.

The creation of a domain-wall brane coupled to gravity is quite straightforward;
we will follow closely the construction in~\cite{Davies:2007tq}.
Beginning with a scalar field $\chi$ and a suitable potential, boundary
conditions are chosen so that $\chi$ develops a kink-like
profile, which can be thought of as a $y$-dependent vacuum expectation
value. As $y\rightarrow\pm\infty$, the value of $\chi$ approaches vacuum and
its energy density rapidly approaches zero.  However, due to the topology
of the vacuum (in general a discrete symmetry is required), a domain wall
forms around $y=0$.  The combination of gradient and vacuum energy in the
core of this object plays an analogous role to the brane tension $\sigma$
in the fundamental case of the previous section.  The shape of the distribution of stress-energy due
to the $y$-dependent profile of $\chi$ is a smooth version of the fundamental
delta-function brane.  In
the non-cosmological case, i.e.  when one seeks the Minkowski
metric on the brane, the solution for the metric then yields
a correspondingly smooth version of the $e^{-\mu |y|}$ warp factor
in~\eqref{eq:rs-metric}.

Because this domain-wall brane is extended in the extra dimension $y$, any
fields that were previously brane-localised by the delta function are no
longer strictly located at $y=0$.  Rather,
such fields (typically the standard model fields) must first be written as
full 5d fields, which are coupled to $\chi$ in such a way as to produce a
Kaluza-Klein tower of 4d fields on the domain wall (in the Minkowski-brane
example see e.g.~\cite{George2007}).  The ground state profile of the 4d
tower has a Gaussian like shape which, when squared\footnote{The squaring
of the extra-dimensional profile comes from the normalisation of the
kinetic term, which is quadratic in the field, hence quadratic in the
profile.}, reduces to a delta-distribution in the limit of an
infinitely-thin domain wall.  

Here we are interested in the more general cosmological case. Our objective is to 
understand the effective 4-dimensional metric on
such a domain-wall brane and how the localised fields propagate in that spacetime. 
In the fundamental-brane case, 4d fields are
located at exactly $y=0$ and have no $y$ degrees of freedom.  The
4d metric they feel is thus just the 5d metric evaluated at
$y=0$, and for the RS metric~\eqref{eq:rs-metric} this slice is
just 4d Minkowski spacetime.  For the cosmological
metric~\eqref{eq:cosmo-metric} the slice at $y=0$ has the form
\begin{equation}
\label{eq:fund-4d}
ds^2 = -n^2(t,y=0)dt^2 + a^2(t,y=0)\gamma_{ij} dx^i dx^j.
\end{equation}
By scaling $t$ such that $n(t,y=0)=1$, it is clear that the
effective 4d metric is of the FRW form, with the effective scale
factor defined by $a_\text{eff}(t)=a_0(t)\equiv a(t,y=0)$.  The
solutions to the 5d Einstein equations given in the previous
section then describe how $a_\text{eff}$ evolves, and hence
describe the spacetime in which the localised fields propagate.
In this fundamental-brane scenario, each field has the same time
(with the same normalisation) and feels the same scale factor, and
so it is sensible to say that the effective 4d metric is unique
and defined by~\eqref{eq:fund-4d}.

For the domain-wall brane scenario things are quite different and,
as we demonstrate explicitly below, we are led to
abandon the question ``what is the effective scale factor on the
brane?'', and allow that \emph{different fields may propagate in different spacetimes}.  
The essential reason for this comes from the extended nature of
the profiles, as the associated fields are now sensitive to the
metric around $y=0$, not just the slice exactly at $y=0$.  Since
the cosmological evolution of the slices in the vicinity of the
brane are misaligned (they expand at different rates), there is
a kind of ``dimensional parallax'' effect, whereby different
species of particle are subject to a different averaging (they
have a different perspective) of the slices.

We note here that for the Minkowski domain-wall brane with a
smoothed-out version of the RS metric~\eqref{eq:rs-metric}, things
are much simpler, because each 4d slice is proportional to
Minkowski spacetime. Therefore, the Minkowski part essentially factorises
out of the averaging integral and each field feels the same
spacetime.

The analysis in Sec.~\ref{sec:fund} determined the effective
scale factor $a_0$ in the case of a general brane-localised source
parameterised by $\rho$ and $p$.  For the
domain-wall brane scenario, we need to look at the sources from
the more fundamental level of classical fields. The general strategy
is to identify the kinetic term in the action for the
relevant field, integrate out the extra dimension $y$, and then to match the
resulting 4d effective action to the canonical 4d action
for such a field in an FRW background.


\subsection{A localised scalar field}
\label{subsec:scalar}

We begin by considering a scalar field, turning to fermions in the next subsection.
We take the 5d metric given by Eq.~\eqref{eq:cosmo-metric} and a 5d scalar field
$\Phi(t,x^i,y)$ separated, for reasons we shall expand on below, as $\Phi(t,x^i,y)=f(t,y)\phi(t,x^i)$.  
The objective is to determine the effective 4d spacetime on which the relevant 4d field $\phi$ propagates.  

The action for a 5d scalar $\Phi(t,x^i,y)$ with metric $g_{MN}$ is
\begin{equation}
\label{eq:act-5d-scalar}
\mcs_5 = \int \! d^4x \! \int \! dy \; \sqrt{g} \left[
        -\frac{1}{2} g^{MN} \partial_M \Phi \partial_N \Phi - U(\Phi)
    \right] \ ,
\end{equation}
where the potential $U$ may contain couplings of $\Phi$ to the
domain-wall field (to localise $\Phi$) or couplings to other fields.
Using the metric ansatz~\eqref{eq:cosmo-metric} we then obtain
\begin{equation}
\mcs_5 = \int \! d^4x \! \int \! dy \; na^3b \sqrt{\gamma} \; \left[
        -\frac{1}{2} \big( -n^{-2} \dot\Phi^2
            + a^{-2} \gamma^{ij} \partial_i \Phi \, \partial_j \Phi 
            + b^{-2} (\partial_y \Phi)^2 \big)
        - U(\Phi)
    \right] \ .
\end{equation}

By analogy with the flat case, our first instinct might be to separate variables by writing 
$\Phi(t,x^i,y) = \sum_n f_n(t,y) \pi_n(x^i)$.  However, in the case of a time-dependent metric such an expansion 
makes it difficult to identify a 4d scalar field, since the time
dependence of, for example, a 4d plane wave, is consumed by the
profile $f_n$, and $\pi$ becomes merely a static spatial wave.  The next obvious suggestion is to instead write 
$\Phi(t,x^i,y) = \sum_n f_n(y) \phi_n(t,x^i)$, so that
$\phi_n$ can be identified as a 4d Kaluza-Klein mode with extra
dimensional profile $f_n$.  Here however, we encounter a different problem, namely that the time variation of the
metric components implies that the extra-dimensional profile
will in general change with time.  

We overcome these obstacles by noting that there are really two time scales in the problem: the cosmological
time scale of the evolution of the background metric, and the time
scale associated with particle physics processes.  With this in mind, we consider the following
separation of variables
\begin{equation}
\Phi(t,x^i,y) = \sum_n f_n(t,y)\phi_n(t,x^i) \ .
\label{sepscales}
\end{equation}
The possible ambiguity in the time dependence (whether it appears in
$f_n$ or $\phi_n$) is resolved by the requirement that $\phi_n$ satisfies
the 4d Euler-Lagrange equation, which will be specified below.  In order
for $\phi_n$ to be identified as a propagating 4d field, it must also carry
the majority of the time dependence, hence we impose the condition
$\dot f_n / f_n \ll \dot \phi_n / \phi_n$.  These requirements formally
identify the class of solutions for $f_n$ that we are allowing.

One should consider this prescription a separation of scales, rather than
a strict separation of variables. Quantitatively, $\dot \phi_n / \phi_n \sim E$ where $E$ is the energy
of the particle, and $\dot f_n / f_n \sim H$ where $H$ is the Hubble
constant.  In natural units we have $H \sim 10^{-32} \text{eV}$,
which is tiny compared to the typical energy of a particle.  In
what follows, we therefore neglect all time derivatives of $f_n$ and of
the metric components $n$, $a$ and $b$, since they are much smaller than
the other terms in the action.  

From now on we focus on a single mode of the KK tower and drop the
subscript $n$.  Then, with the prescription~(\ref{sepscales}) and the
assumptions regarding small time-derivatives, the action becomes
\begin{equation}
\mcs_5 = \int \! d^4x \sqrt{\gamma} \int \! dy \left[
        -\frac{1}{2} \left( -\frac{a^3b}{n} f^2 \dot \phi^2
            + nab f^2 \gamma^{ij} \partial_i \phi \, \partial_j \phi
            + \frac{na^3}{b} f'^2 \phi^2 \right)
        - na^3b U
    \right] \ ,
\end{equation}
where a prime denotes a derivative with respect to $y$.  The third term, proportional
to $\phi^2$, will contribute to the potential $U$.  Integrating
over the extra dimension yields the 4d effective action
\begin{equation}
\label{eq:scalar-eff-act}
\mcs_4 = \int \! d^4x \sqrt{\gamma} \left[
    -\frac{1}{2} \left( -F(t) \dot \phi^2
        + G(t) \gamma^{ij} \partial_i \phi \partial_j \phi \right) + \ldots \right] \ ,
\end{equation}
where we have written only the kinetic terms explicitly, and defined
\begin{align}
\label{eq:fg-defn}
F(t) &\equiv \int f^2 \frac{a^3b}{n} dy, & G(t) &\equiv \int f^2 nab \; dy.
\end{align}

The action~\eqref{eq:scalar-eff-act} is almost what we are looking for, but what remains is to
correctly identify the 4d line element describing the spacetime within which $\phi$ propagates.  To do
this, we match to the
prototype line element
\begin{equation}
\label{eq:metric-4d}
ds_4^2 = -T^2(t) dt^2 + X^2(t) \gamma_{ij} dx^i dx^j \ ,
\end{equation}
and the corresponding prototype action
\begin{equation}
\label{eq:scalar-act}
\mcs_4^\text{(proto)} = \int \! d^4x T(t)X^3(t) \sqrt{\gamma} \; \left[
        -\frac{1}{2} \left( -T^{-2}(t) \dot \phi^2
            + X^{-2}(t) \gamma^{ij} \partial_i \phi \partial_j \phi \right)
    \right] \ .
\end{equation}
Matching the effective action~\eqref{eq:scalar-eff-act} with the 4d
prototype~\eqref{eq:scalar-act} we then obtain
\begin{equation}
F(t) = T^{-1}(t) X^3(t)\ , \ \ \ \  G(t) = T(t) X(t) \ .
\end{equation}
Solving for $T(t)$ and $X(t)$ gives
\begin{align}
\label{eq:scalar-t}
T(t) &= F^{-1/4}(t) G^{3/4}(t) = \left(\int f^2 \frac{a^3b}{n} \, dy \right)^{-1/4} \left(\int f^2 nab \, dy \right)^{3/4}, \\
\label{eq:scalar-x}
X(t) &= F^{1/4}(t) G^{1/4}(t) = \left(\int f^2 \frac{a^3b}{n} \, dy \right)^{1/4} \left(\int f^2 nab \, dy \right)^{1/4} \ .
\end{align}

The time-dependent functions $T(t)$ and $X(t)$ define, along with~\eqref{eq:metric-4d}, the effective 4d line
element followed by the field $\phi$.  As we shall soon demonstrate,
we are free to rescale $f$ by an arbitrary (slowly varying)
function of time, and we can use this freedom to fix $T=1$.
This corresponds to choosing a canonical time coordinate.  The 
scale factor for $\phi$ is then precisely
\begin{equation}
\label{eq:scalar-aeff}
a_\phi(t) = X(t) \ .
\end{equation}
Notice
that the temporal behaviour of $X(t)$ (and hence $a_\phi$) is inherited from
the time-dependence of the metric components and possibly $f(t,y)$,
all of which are taken to be slowly varying.

This result for the effective scale factor immediately raises two important points.
The first is that scalar modes with different profile functions will have
different definitions of the scale factor $a_\phi$. Thus, it is not possible to
define a unique scale factor for this 4d effective theory.  Instead, each 4d scalar
field, whether it arises from a different 5d field, or is merely a different KK
mode of the same 5d field, propagates according to a different effective 4d metric.

The second interesting point is that the procedure above will yield a different
result for a fermionic field (and also other spin fields) due to the difference
arising from the spin connection in the
kinetic term.  We will follow this point up in the next section where we explicitly perform the relevant calculation for a fermion.

As a consistency check, we consider Eqs.~\eqref{eq:scalar-t} and~\eqref{eq:scalar-x}
in the limit of an infinitely-thin domain wall.  In such a limit, the square of a typical
ground state profile $f$ becomes proportional to a delta-function distribution,
$f^2\rightarrow\delta(by)$.  This comes from the kinetic term for $\phi$, which is
quadratic in $f$, and must be normalised such that, in the thin brane limit,
$\smallint f^2 d(by)=1$.
The integrals for $T(t)$ and $X(t)$ can then be performed analytically yielding
$T(t)=n(t,y=0)$ and $X(t)=a(t,y=0)$.  These coincide with the fundamental brane case, in
which the 4d metric is the 5d metric evaluated on the brane.

A further check on our derivation can be made by looking at the separable (but less
general) version of the cosmological metric, given by
\begin{equation}
\label{eq:sep-cosmo-metric}
ds_5^2 = c^2(y) [-dt^2 + \hat a^2(t) \gamma_{ij} dx^i dx^j] + \hat b^2(y) dy^2 \ .
\end{equation}
This ansatz allows for AdS$_4$ and dS$_4$ brane solutions, as detailed
in~\cite{DeWolfe:1999cp,Karch2000,Kaloper:1999sm,Gremm:2000dj,Afonso:2006gi,Flanagan:2001dy,Davidson:2000hi,Slatyer2007}.
The effective 4d metric for $\phi$ then has
$T(t)=(\smallint f^2 c^2 \hat b \; dy)^{1/2}$ and $X(t)=\hat a(t) (\smallint f^2 c^2 \hat b \; dy)^{1/2}$.
Note that $T$ is constant ($f$ will be time-independent; see later) and we can
normalise $f$ to make $T=1$, and then find that $X(t)=\hat a(t)$.  In this case we
again recover the known result, namely that all fields on the brane feel the same
metric.

To complete this formal analysis of $\Phi$ we determine the differential equation
satisfied by the profile function $f(t,y)$.  Our definitions above for the separation
of scales ensure that $\phi$ behaves as a 4d scalar field in a spacetime
characterised by $T(t)$ and $X(t)$.  This means that $\phi$ will satisfy the Euler-Lagrange
equation
\begin{equation}
\label{eq:el-4d}
-\frac{1}{T^2} \ddot \phi
    + \frac{1}{X^2} \gamma^{ij} \left( \partial_i \partial_j \phi
        - \Gamma^{(3)\,k}_{\phantom{(3)}\,ij} \partial_k \phi \right) = m^2 \phi \ ,
\end{equation}
where $\Gamma^{(3)\,k}_{\phantom{(3)}\,ij}$ are the connection
coefficients associated with the 3-space metric $\gamma_{ij}$
and $m$ is the effective 4d mass of $\phi$.  The parenthesised
term on the left hand side is simply the double
covariant-derivative of $\phi$ with respect to $\gamma_{ij}$.
Note that we are ignoring time derivatives of $T(t)$ and $X(t)$,
which are much smaller that the derivatives of $\phi$.

Now consider the 5d Euler-Lagrange equation for $\Phi$
\begin{equation}
\label{eq:el-5d}
g^{MN} \left( \partial_M \partial_N \Phi - \Gamma^{(5)\,P}_{\phantom{(5)}\,MN} \partial_P \Phi \right)
    = \frac{\partial U}{\partial \Phi} \ ,
\end{equation}
where $\Gamma^{(5)\,P}_{\phantom{(5)}\,MN}$ are the 5d connection
coefficients. We first separate variables, neglect time derivatives of $n$, $a$, $b$ and
$f$, and use Eq.~\eqref{eq:el-4d} to eliminate the spatial
derivatives of $\phi$ (thus the 4d mass will appear).  We then 
linearise the equation, yielding
\begin{equation}
\label{eq:f-de-prelim}
\left[ f''
    + \left( \frac{n'}{n}+\frac{3a'}{a}-\frac{b'}{b} \right) f'
    + b^2 \left( m^2 \frac{X^2}{a^2} - U^{(1)} \right) f \right] \phi
    + \frac{b^2}{a^2} \left( \frac{X^2}{T^2} - \frac{a^2}{n^2} \right) f \ddot\phi = 0 \ ,
\end{equation}
where $\partial U / \partial\Phi = U^{(1)} \Phi + \mco(\Phi^2)$.
Notice the appearance of the $\ddot\phi$ term, which is absent when we specialise
to Minkowski spacetime on the brane.  There are two reasons for this. First,
there is a mismatch between the 5d ratio of the time and $3$-space metric
factors, and the corresponding effective 4d ratio; $a^2/n^2 \ne X^2/T^2$.  For a
Minkowski brane these ratios are equal because each 4d slice of the 5d metric is
proportional to Minkowski spacetime.  Second, we have employed a separation of
scales rather than the usual separation of variables (which did not work in this setting).
This $\ddot\phi$ term then quantifies the \emph{inability of the domain-wall brane to
localise proper 4d effective fields}, at least in a cosmological background.

To proceed, we need to eliminate the $\phi$ and $\ddot\phi$ factors so that we
have an equation that can, at least in principle, be used to solve for $f$.  To
this end we solve the 4d Euler-Lagrange equation~\eqref{eq:el-4d} (in flat
space; $k=0$) and find ``plane waves'' of the form
\begin{equation}
\phi(t,x^i) \propto  \exp(-i T^2Et + i X^2 \gamma_{ij} p^i x^j) \ ,
\end{equation}
where $E$ is the energy of the wave, $p^i$ is its momentum, and
$T^2E^2 = X^2 \gamma_{ij} p^i p^j + m^2$.  Then $\ddot\phi=-E^2T^4\phi$,
and equation~\eqref{eq:f-de-prelim} becomes
\begin{equation}
\label{eq:f-de}
f''
    + \left( \frac{n'}{n}+\frac{3a'}{a}-\frac{b'}{b} \right) f'
    + b^2 \left[ m^2 \frac{X^2}{a^2} - U^{(1)}
        - \frac{E^2 T^4}{a^2} \left( \frac{X^2}{T^2} - \frac{a^2}{n^2} \right) \right] f = 0 \ .
\end{equation}
Usually, such an equation depends only on $m$, implying that
although different masses in the KK tower of 4d fields ($\phi_0$, $\phi_1$,
etc.) have different profiles, these profiles are independent of the energy. Here however, 
the equation also depends
on $E$, so that quanta with the same mass but different
\emph{energies} (or momenta) have different profiles.
On the surface, Eq.~\eqref{eq:f-de} looks linear and homogeneous
in $f$, but it is in fact a non-linear integro-differential equation,
since both $T(t)$ and $X(t)$ are defined in terms of $f$.  Nevertheless,
this equation still has the property that $f$ can be rescaled by
a $y$-independent factor, so long as the eigenvalues $m$ and $E$ are
also appropriately rescaled to compensate for the change in $T(t)$ and
$X(t)$.  In fact, since $\dot f \ll E$, as discussed previously, we may even 
take this factor to have a (mild) time-dependence. 
As we advertised earlier, the rescaling of $f$ can be used to choose
a canonical time coordinate, corresponding to fixing $T=1$, which is
achievable precisely because $T(t)$ depends on $f$.

As a check on our derivation, for the separable cosmological
metric~\eqref{eq:sep-cosmo-metric},
the factor $X^2/T^2 - a^2/n^2$ vanishes, and Eq.~\eqref{eq:f-de}
simplifies to the known result (see~\cite{Davies:2007tq})
\begin{equation}
f''
    + \left( \frac{4c'}{c}-\frac{\hat b'}{\hat b} \right) f'
    + \hat b^2 \left[ m^2 \frac{1}{c^2} - U^{(1)} \right] f = 0 \ .
\end{equation}
Note the lack of time dependence, implying that $f$ is a function
of $y$ only.  The profile also no longer depends on the energy of the
mode, just its mass, as usual.

Let us summarise our results for scalar fields.  Given a 5d
background metric, described by functions $n(t,y)$, $a(t,y)$, $b(t,y)$,
and a generic coupling potential $U(\Phi)$,
we may solve Eq.~\eqref{eq:f-de} for $f(t,y)$.  The particular solution
depends on a mass eigenvalue $m$ and an energy $E$.  We may then use
this solution $f(t,y)$ to determine $T(t)$ and $X(t)$ through
Eqs.~\eqref{eq:scalar-t} and~\eqref{eq:scalar-x}.  What
results is the 4d spacetime (described by $T(t)$ and $X(t)$) on which a 4d
quantum field $\phi$ with mass $m$ and energy $E$ propagates.  We are free 
to rescale $f(t,y)$ to impose $T=1$, so that $X(t)$ can then be interpreted as the
effective FRW scale factor.  The crucial result to note is that
\emph{the scale factor depends on the type of field, its coupling
potential, and its 4d mass and momentum}.


\subsection{Localised fermions}
\label{subsec:fermions}

We now turn to fermions, and perform an analogous calculation to determine
the effective scale factor describing the 4-dimensional spacetime on which a
localised fermion field propagates.

For a 5d fermion $\Psi(t,x^i,y)$, the
action is
\begin{equation}
\mcs_{5,\Psi} = \int \! d^4x \! \int \! dy \; \sqrt{g} \; \left[
    \overline\Psi \Gamma^A E_A^{\phantom{A}M} (\partial_M + \omega_M) \Psi
    - U_\Psi \overline\Psi \Psi \right] \ ,
\end{equation}
where $\Gamma^A$ are the 5d flat-space gamma matrices, $E_A^{\phantom{A}M}$ are
the vielbeins and $\omega_M$ is the spin connection\footnote{We
are using $A$, $B$ to denote 5d flat-space indices and $M$, $N$ to
denote 5d curved-space indices.}. The gamma-matrices obey
$\{\Gamma^A, \Gamma^B\}=2\eta^{AB}$, with
$\eta^{AB}=\diag(-1,1,1,1,1)$.  The coefficient $U_\Psi$ of the
mass term will in general be a function of other fields, to allow,
for example, coupling of the fermion to the domain wall.  

As for a scalar field, we perform a separation of scales in time and
separation of variables in space\footnote{There is a subtlety here: we
are assuming that all four components of the Dirac spinor $\psi$ have
the same profile $u$, which may not be warranted.  We expand on this later.}
\begin{equation}
\Psi(t,x^i,y)=u(t,y)\psi(t,x^i)
\end{equation}
and expand the kinetic
terms, ignoring $\dot u$.  The action becomes
\begin{equation}
\mcs_{5,\Psi} = \int \! d^4x \! \int \! dy \; na^3b \sqrt{\gamma} \;
    \left[ u^2 \, \overline\psi \left( -n^{-1} \gamma^0 \dot\psi
            + a^{-1} \gamma^a e_a^{\phantom{a}j} \partial_j \psi \right)
        + \ldots \right] \ ,
\end{equation}
where $\gamma^0$, $\gamma^a$ are the 4d flat-space gamma matrices
with $\{\gamma^\alpha,\gamma^\beta\}=2\eta^{\alpha\beta}$ and
$e_a^{\phantom{a}j}$ are vielbeins for the 3d space, with
$\gamma_{ij}$ the metric\footnote{$\alpha,\beta=0,1,2,3$ are the
4d flat-spacetime indices, $a=1,2,3$ is a 3d flat-space index.}.
As before, we require that the correct powers of $n(t,y)$, $a(t,y)$ and
$b(t,y)$ match with the relevant terms in the prototype 4d fermion action
\begin{equation}
\mcs_{4,\psi}^\text{(proto)} = \int \! d^4x T_\psi (t) X_\psi^3 (t) \sqrt{\gamma} \;
    \overline\psi \left( -T_\psi^{-1}(t) \gamma^0 \dot\psi
        + X_\psi^{-1}(t) \gamma^a e_a^{\phantom{a}j} \partial_j \psi \right) \ ,
\end{equation}
where we have used the prototype line element
$ds_4^2=-T_\psi^2(t)dt^2 + X_\psi^2(t)\gamma_{ij}dx^idx^j$.  Matching kinetic coefficients we obtain
\begin{align}
F_\psi (t) &= \int u^2 a^3b \, dy = X_\psi^3(t), \\
G_\psi (t) &= \int u^2 na^2b \, dy = T_\psi(t) X_\psi^2(t) \ ,
\end{align}
which may be inverted to give
\begin{align}
T_\psi (t) &= F_\psi^{-2/3}(t) G_\psi(t) = \left( \int u^2 a^3b \, dy \right)^{-2/3} \left( \int u^2 na^2b \, dy \right), \\
\label{eq:ferm-x}
X_\psi (t) &= F_\psi^{1/3}(t) = \left( \int u^2 a^3b \, dy \right)^{1/3} \ .
\end{align}

These results are similar to the scalar case. As there, we can rescale $u(t,y)$ by
a slowly varying function of time to enforce $T_\psi=1$, so that
$\psi$ describes a 4d fermion field in a spacetime with effective
scale factor $a_\psi(t)=X(t)$.  Again, the definition of the
effective scale factor depends on the profile of the particular
KK mode $\psi$ that one is interested in.  As before, in the
infinitely-thin domain-wall limit $u^2 \rightarrow \delta(by)$ and
$T_\psi (t) \rightarrow n(t,y=0)$, $X_\psi (t) \rightarrow a(t,y=0)$, which
recovers the known delta-function brane result.  For the separable cosmological
metric~\eqref{eq:sep-cosmo-metric}, $T_\psi$ is a constant and after
normalising $u$ such that $T_\psi=1$ we have $X_\psi(t)=\hat a(t)$:
the standard result.

To identify the equation satisfied by $u(t,y)$, we impose the
requirement that $\psi$ satisfies the 4d
Euler-Lagrange equation with mass $m_\psi$ and use this to
eliminate the spatial derivatives of $\psi$ from the 5d Euler-Lagrange equation for $\Psi$.
This yields
\begin{equation}
\label{eq:u-de}
\left[ u' + \left( \frac{n'}{2n} + \frac{3a'}{2a} \right) u \right] \gamma^5 \psi
    + b \left(m_\psi \frac{X}{a} - U_\Psi \right) u \psi
    - \frac{b}{a} \left( \frac{X_\psi}{T_\psi} - \frac{a}{n} \right) u \gamma^0 \dot\psi = 0 \ ,
\end{equation}
where $\gamma^5=i\gamma^0\gamma^1\gamma^2\gamma^3$.  This has a
similar structure to the scalar version~\eqref{eq:f-de-prelim}; in particular the $\dot\psi$ term
quantifies the deviation from $\psi$ being a 4d field in the usual, infinitely-thin brane definition.

The appearance of $\gamma^0$ and $\gamma^5$ in
Eq.~\eqref{eq:u-de} means localised states on the domain wall have an unusual Dirac structure.  
In the Minkowski-brane case, the $\gamma^0\dot\phi$ term is absent and this leads to the
localisation of chiral states, which are eigenspinors of
$\gamma^5$.  With the presence of $\gamma^0$, one would na\"ively
seek Dirac states which are eigenspinors of both
$\gamma^0$ and $\gamma^5$, which is impossible!  It therefore seems that
the time-dependent background metric leads to unconventionally
localised spinor states.  To understand this problem more deeply,
consider seeking solutions to Eq.~\eqref{eq:u-de}
when $\psi=\psi_L$ (and $u=u_L$) is left-chiral, i.e. $\gamma^5\psi_L=-\psi_L$ and
$m_{\psi_L}=0$.  Using a plane wave solution for $\psi_L$ and
expanding its Weyl components in order to evaluate
$\gamma^0\dot\psi_L$, we obtain two independent equations for
$u_L$
\begin{align}
\label{eq:ul-de-1}
u_L' + \left( \frac{n'}{2n} + \frac{3a'}{2a} \right) u_L + b U_\Psi u_L &= 0, \\
\label{eq:ul-de-2}
\frac{b}{a} \left( \frac{X_{\psi_L}}{T_{\psi_L}} - \frac{a}{n} \right) u_L E_{\psi_L} T_{\psi_L}^2 &= 0 \ ,
\end{align}
where $E_{\psi_L}$ is the energy of the chiral plane-wave spinor
$\psi_L$.  

With a non-trivial background, the only solution to
Eq.~\eqref{eq:ul-de-2} is $u_L=0$. Thus, there are no localised
left-chiral spinors.  It may be possible to rectify this problem
and find localised states which have a certain definite spinor
structure by relaxing the separation ansatz (recall that each
component of the Dirac spinor was assumed to have the same profile
$u$, which may be an overly strict assumption), but we will not
pursue this line of thought further here.


\subsection{The effective Newton's constant}

We shall now briefly discuss how to determine the effective
Planck mass, and hence Newton's constant, describing the
strength of gravity on the brane\footnote{We are concerned here with
the definition of the Planck mass for use in cosmological situations,
e.g. in Eq.~\eqref{eq:eff-friedmann-3}, as opposed to its use in
Cavendish-like experiments.}.  Usually, one expands the 5d
Ricci scalar in the Einstein-Hilbert action in terms of its 4d
counterpart, and the numerical pre-factor is identified as the
Planck mass.  For example, in the RS2 model, one uses the metric
\begin{equation}
ds_5^2 = e^{-2\mu |y|} g^{(4)}_{\mu \nu}(x^\mu) dx^\mu dx^\nu + dy^2 \ ,
\end{equation}
for which the Einstein-Hilbert action can be written as
\begin{align}
\mcs_\text{EH} &= \int d^4x \int dy \sqrt{-g} \; M_5^3 R \\
    &\supset \int d^4x \int dy \; e^{-4\mu |y|} \sqrt{-g^{(4)}} \; M_5^3 \; e^{2\mu |y|} R^{(4)} \ ,
\end{align}
where $R^{(4)}$ is the
4d Ricci scalar associated with $g^{(4)}_{\mu \nu}$. One then identifies
the effective 4d Planck mass as
\begin{equation}
M_P^2 \equiv M_5^3 \int_{-\infty}^{\infty} e^{-2\mu |y|} dy
    = \frac{M_5^3}{\mu} \ ,
\end{equation}
which agrees with the result obtained from the effective Friedmann
equation in the fundamental brane scenario, Eq.~\eqref{eq:mp}.


Following this approach for domain-wall cosmology, we begin by considering
how the 4d Ricci scalar is embedded in the 5d one. However, there is a 
problem with this approach, because the metric factors for $t$ and $x^i$
behave differently at the 5d level, and so the 5d Ricci scalar does not
separate into a 4d piece plus other terms.  To make progress, one might consider restricting attention
to the 3d Ricci scalar (constructed from the 3d spatial metric $\gamma_{ij}$), which does separate, and
identifying its pre-factor in the Einstein-Hilbert action as the Planck mass.
In other words, we consider just the three spatial components of the metric
perturbations to determine the gravitational coupling, instead of using
the temporal and spatial components together.  Explicitly, we
first write the 5d metric in the general form
\begin{equation}
\label{eq:metric-5d-pert}
ds_5^2 = -n^2(t,y) dt^2 + a^2(t,y) \xi_{ij}(t,x^i) dx^i dx^j + b^2(t,y) dy^2 \ .
\end{equation}
The 5d Einstein-Hilbert action can then be expanded as
\begin{align}
\label{eq:eh-5d}
\mcs_\text{EH} &= \int d^4x \int dy \sqrt{-g} \; M_5^3 R \\
\label{eq:eh-5d-expand}
    &\supset \int d^4x \int dy \; na^3b \sqrt{\xi} \; M_5^3 \; a^{-2} R^{(3)} \ ,
\end{align}
where $R^{(3)}$ is the Ricci scalar constructed from $\xi_{ij}$.  
The goal is to match this to the 4d prototype Einstein-Hilbert action
associated with the general prototype metric 
\begin{equation}
ds_4^2=-T_M^2(t)dt^2 + X_M^2(t) \xi_{ij}(t,x^i) dx^i dx^j \ ,
\end{equation}
which yields
\begin{align}
\mcs_\text{EH}^\text{(proto)} &= \int d^4x \sqrt{-g^{(4)}} \; M_P^2 R^{(4)} \\
\label{eq:eh-4d-expand}
    &\supset \int d^4x T_M(t) X_M^3(t) \sqrt{\xi} \; M_P^2 \; X_M^{-2}(t) R^{(3)} \ .
\end{align}

By comparing Eqs.~\eqref{eq:eh-5d-expand} and~\eqref{eq:eh-4d-expand},
we can infer that the 5d theory produces 3d (three spatial) metric
perturbations with effective Planck mass
\begin{equation}
\label{eq:eff-mp}
M_P^2 \equiv \frac{M_5^3}{T_M(t)X_M(t)} \int nab \; dy \ .
\end{equation}
This result requires us to specify the 4d spacetime
(by specifying $T_M(t)$ and $X_M(t)$) before we can know the
Planck mass.  As shown in the previous sub-sections, the 4d
spacetime is dependent on the particle species, and so we
obtain a \emph{species dependent Planck mass}.  This may not be
so surprising given that each species follows a different line
element, but it is also possible that the assumption of matching
only the 3d Ricci scalar is unwarranted.

Perhaps a more sophisticated calculation would try to elucidate
the effective 4d Einstein equation, or at least the leading order
contribution.  Ultimately, we would like to identify  $1/2M_P^2$ as the constant of proportionality
between the dominant (first order) contributions to the 4d Einstein tensor
$^{(1)}G^{(4)}_{\mu\nu}$ and the
stress-energy tensor $^{(1)}T^{(4)}_{\mu\nu}$ for a given field in the
thin, large-tension brane limit.  One possible way to perform this
calculation would be to analyse the equations of motion for metric
perturbations.  In the case of fundamental-brane cosmology, much
of the ground-work for such an analysis has been performed; see
for example~\cite{Langlois:2000ia,Langlois:2000ns,Easther2003}.
For a domain-wall brane, extra complications arise, again due to
the averaging of the metric over the extra dimension.  Further, it
seems that in order to identify the 4d metric perturbations, one
is forced to perform a separation of scales, as in the scalar and
fermion case.  Such a calculation is beyond the scope of
this paper, and for our purposes here we adopt Eq.~\eqref{eq:eff-mp} as an
approximate definition of the effective Planck mass.


\section{Effective scale factor for a thin domain wall}
\label{sec:thinwall}

Having developed the general framework for a domain wall with localised matter fields and the associated 
four-dimensional metric, we would like to better understand the behaviour of
the effective scale factors $a_\phi$ and $a_\psi$.  These will, of
course, depend on the details of the domain wall construction. Furthermore, we need to
solve explicitly for the
metric components $n(t,y)$, $a(t,y)$ and $b(t,y)$ in
the presence of this domain wall.  We are unable to find
analytic solutions for a coupled domain-wall gravity system, and
numerical solutions are beyond the scope of
this initial work.  To make progress therefore, we will assume that the
domain wall is extremely thin and that therefore the solutions for the
metric components are well approximated by the set of
equations~\eqref{eq:metric-bulk}.  

The profiles of the domain-wall fields will play a role in determining the
profiles $f(t,y)$ and $u(t,y)$ of the localised scalars and fermions
respectively.  These localised fields will then contribute to the
total stress-energy and this back-reaction will modify the metric components.  
However, here we shall ignore such back-reaction effects and consider the thin domain wall as 
a small perturbation to the fundamental-brane scenario presented in
Sec.~\ref{sec:fund}.  In order for this perturbative approach to work,
it is necessary that the brane localised sources be relatively small,
meaning that $\epsilon\ll 1$.

To compute $a_\phi$ within this approximation scheme, we 
first normalise $f(t,y)$ such that $T=1$, by defining
\begin{equation}
f(t,y)=\tau(t)\tilde f(t,y)
\end{equation}
so that
\begin{equation}
T(t)= \tau(t) \, \tilde F^{-1/4}(t) \, \tilde G^{3/4}(t) \ , \ \ \ \ \  X(t) = \tau(t) \, \tilde F^{1/4}(t) \, \tilde G^{1/4}(t) \ ,
\end{equation}
where $\tilde F(t)$ and $\tilde G(t)$ are defined as in
Eq.~\eqref{eq:fg-defn} but with $f(t,y)$ replaced by $\tilde f(t,y)$.
Enforcing $T=1$ gives $\tau (t)=\tilde F^{1/4}(t) \, \tilde G^{-3/4}(t)$ so that 
$X(t)=\tilde F^{1/2}(t) \, \tilde G^{-1/2}(t)$.  We may then compute
$\tilde F(t)$ and $\tilde G(t)$ by substituting in for the bulk metric
solutions~\eqref{eq:metric-bulk} yielding, for example,
\begin{equation}
\tilde G(t) = \int \tilde f \, an \; dy = a_0 \int \tilde f \left[ e^{-2\mu|y|} -
        (\epsilon+\tilde\epsilon) e^{-\mu|y|} \sinh (\mu|y|)
        + \epsilon\,\tilde\epsilon \, \sinh^2 (\mu|y|) \right] dy \ .
\end{equation}
Requiring that the localisation profile $\tilde f(t,y)$ be sharply peaked at the
centre of the domain wall ($y=0$) and fall off rapidly in the bulk translates to
$\tilde f^2(t,y) \sinh^2 (\mu|y|) \rightarrow 0$ as $y\rightarrow \pm \infty$.  
This condition is consistent with the sufficiently-thin domain-wall brane we are dealing with here.  
Thus, we may
ignore the second order term ${\cal O}(\epsilon\,\tilde\epsilon)$ and write
\begin{equation}
\tilde G(t) = a_0 \left[ I_1(t) - (\epsilon+\tilde\epsilon) I_2(t) \right] \ ,
\end{equation}
where
\begin{align}
I_1(t) &= \int \tilde f^2(t,y) e^{-2\mu|y|} dy \ , \\
I_2(t) &= \int \tilde f^2(t,y) e^{-\mu|y|} \sinh (\mu|y|) dy \ .
\end{align}
These integrals, $I_1(t)$ and $I_2(t)$, are dependent on the exact
form of the extra-dimensional profile $\tilde f(t,y)$. However, if the profile is
sufficiently peaked, as we are assuming, we have
$I_2(t) \ll I_1(t)$, because $\sinh (\mu|y|) \sim 0$ close to the centre
of the domain wall.  Under these assumptions, we may compute
\begin{equation}
\tilde F(t) = a_0^3 \left[ I_1(t) - (3\epsilon-\tilde\epsilon) I_2(t) \right] \ ,
\end{equation}
so that the effective scale factor for the scalar field becomes
\begin{align}
a_\phi(t) = X(t) &= a_0(t) \left[ 1 - (\epsilon-\tilde\epsilon)\frac{I_2(t)}{I_1(t)} \right] \\
\label{eq:thinscalarscale}
    &= a_0(t) \left[ 1 + \frac{\dot\epsilon}{H_0}\frac{I_2(t)}{I_1(t)} \right] \ .
\end{align}
This is one of the main results of our paper -- an explicit, quantitative computation of the
corrections to the effective 4-dimensional scale factor arising from considering a
domain-wall brane, rather than a fundamental one.  The corrections are proportional to the
ratio between the rate of change of energy density on the brane and the brane tension, and inversely
proportional to the effective Hubble parameter.  The corrections also depend in a
non-trivial way on the specific localisation profile of the associated field, so that
different fields are corrected differently.  

The expression~\eqref{eq:thinscalarscale} satisfies $a_\phi \rightarrow a_0$ for the
independent limits of a Minkowski brane with no sources
($\epsilon \rightarrow 0$), and an infinitely-thin brane
($I_2 \rightarrow 0$).  For a concrete example of this latter limit,
consider the profile\footnote{The time dependence of this sample $\tilde f$
would arise from the time dependence of the parameter $w$, corresponding
to the brane thickness changing over time.}
\begin{equation}
\tilde f^2(t,y) = \frac{\Gamma(w+\tfrac{1}{2})}{\sqrt{\pi}\Gamma(w)} \mu [\cosh(\mu y)]^{-2w} \ ,
\end{equation}
which obeys $\tilde f^2 \rightarrow \delta(y)$ as
$w \rightarrow \infty$ (the thin domain-wall limit) and is a typical example
of smooth localisation factors (see, for example,~\cite{George2007}).  It is then straightforward to
compute
\begin{equation}
\frac{I_2(t)}{I_1(t)} =
    \frac{2\Gamma(w+\tfrac{1}{2}) - \sqrt{\pi}\Gamma(w)}
        {2\sqrt{\pi}\Gamma(w+1) - 4\Gamma(w+\tfrac{1}{2})}
    \quad \underrightarrow{w\rightarrow\infty} \quad \frac{1}{\sqrt{\pi w}} \ ,
\end{equation}
which vanishes in the infinitely-thin wall limit.  A better
approximation for this quantity can be found by solving the
differential equation~\eqref{eq:f-de} for $\tilde f(t,y)$, using the
background metric components $n(t,y)$ and $a(t,y)$.

For a fermion field the result for the effective scale factor is
almost identical to the scalar case
\begin{equation}
\label{eq:thinfermscale}
a_\psi(t) = a_0(t) \left[ 1 + \frac{\dot\epsilon}{H_0}\frac{J_2(t)}{J_1(t)} \right] \ ,
\end{equation}
where the relevant integrals are
\begin{align}
J_1(t) &= \int \tilde u^2(t,y) e^{-3\mu|y|} dy, \\
J_2(t) &= \int \tilde u^2(t,y) e^{-2\mu|y|} \sinh \mu|y| dy \ ,
\end{align}
and $\tilde u(t,y)$ is defined in a similar way to $\tilde f(t,y)$.

The results from this section, namely Eqs.~\eqref{eq:thinscalarscale}
and~\eqref{eq:thinfermscale}, are concrete expressions for
modifications to cosmology in a domain-wall brane construction, and are
the starting point for an analysis of the constraints on such
theories from observations. We expect that a species-dependent scale
factor should have an impact on a vast array of cosmological
observables, including the predictions of BBN, the era of recombination and the
spectra of the microwave background and large scale
structure. Acceptable cosmological behaviour should imply constraints on
the brane tension $\sigma$, which appears in
$\epsilon$ and $\mu$, and the width of the domain wall, which
enters implicitly through the localisation profiles $\tilde f$
and $\tilde u$.  


\section{Conclusions}

A self-contained and self-consistent construction of a
five-dimensional theory requires that the four-dimensional brane on
which the standard model fields are confined be formed as a domain
wall. Such an object has finite thickness, and couples in a
variety of ways to the particle physics and gravitational fields
of the theory.

Dimensional reduction from five down to four dimensions involves
integrating over the extra dimension in the 5-dimensional action.
Each term in the resulting 4-dimensional effective action then
contains numerical factors arising from these integrals over the
extra-dimensional profiles of the fields.  These factors must be
absorbed into the fields in order to normalise the kinetic terms.
In the case of a Minkowski metric on the brane, this absorption
can be carried out via a single definition, to obtain an effective
field theory applicable to all fields. 

In this paper we have considered the problem of cosmology on a
4-dimensional domain-wall brane. In contrast to the procedure of
obtaining a Minkowski metric, this is a complicated process and,
in particular, during the normalisation of the kinetic terms, one
must take into account the fact that \emph{different fields may
feel different spacetimes}.  Thus, for domain-wall branes it is
not sensible to ask the question ``what is the effective scale
factor on the brane?''.  Since the brane has non-trivial dependence
on the extra-dimensional coordinate $y$, and since the metric
components $n(t,y)$ and $a(t,y)$ are not proportional to each other
in the bulk, each 4-dimensional slice at constant $y$ corresponds
to a different spacetime.  The effective 4-dimensional spacetime
for a localised field with a smooth profile in $y$ will thus be an
average over all the different slices.  This produces a kind of
``dimensional parallax'' effect, since different fields have
different averages, and thus a different ``perspective'' of the
cosmological evolution of each slice.

Therefore, rather than seeking the effective scale factor on the
brane (which was possible in the fundamental brane case), we must
instead ask: ``what is the effective 4d spacetime in which a given
localised field propagates?''.  For each low-energy 4-dimensional field
(each species and each mode of the KK tower), we may answer this
question by determining the effective 4-dimensional line element $ds_4^2$.
If this line element takes the form of an FRW line element, then
we may define an effective scale factor for the associated field.
This is as close as we are able to come to answering our original
question.

In the case of a localised scalar field, the effective
scale factor is given in general by Eqs.~\eqref{eq:scalar-x}
and~\eqref{eq:scalar-aeff}, and for fermions one
obtains~\eqref{eq:ferm-x}.  For the case of a domain-wall brane which
is thin enough such that one can approximate the metric components
$n(t,y)$, $a(t,y)$ and $b(t,y)$ with the solutions from the
fundamental-brane scenario, the effective scale factors for scalars
and fermions are~\eqref{eq:thinscalarscale} and~\eqref{eq:thinfermscale}
respectively.  All of these results reduce, in the infinitely-thin
domain-wall limit, to the results obtained for fundamental brane
cosmology, where the effective scale factor is the bulk metric
component $a(t,y)$ evaluated at $y=0$.

Beyond this basic difference between the fundamental-brane and the
domain-wall case, there are a number of other interesting
consequences of our construction. In the cosmological scenario,
the non-trivial averaging of the metric over $y$ means that, just
as different modes of a KK tower have different extra-dimensional
profiles, it is also true that different \emph{energies} of the
same mass have different profiles.  Thus, particles with different
energies feel a different scale factor!  An outstanding problem is
that of defining a unique Planck mass (if possible) at the 4d level.
We have provided some insight into the solution to this problem,
in the form of the initial approximation~\eqref{eq:eff-mp}, but
a full treatment is beyond the scope of this paper.

The novel features we have presented -- species-dependent
scale factors and non-unique Planck masses -- will lead to
potentially observable cosmological phenomena.  We are currently
performing a detailed phenomenological analysis to determine how
these effects constrain parameters such as the width and tension
of the domain wall, and the extent to which they may allow new
approaches to cosmological problems.


\acknowledgments

DPG and RRV thank Archil Kobakhidze for discussion and comments on the
manuscript.
DPG also thanks the Department of Physics at Syracuse University for their
warm hospitality, where, during his visit, he had the opportunity to carry
out the majority of this work.
MT would like to thank the University of Melbourne
for the award of a Sir Thomas Lyle Fellowship, during which this work
began, and the School of Physics at Melbourne for their hospitality during
his tenure there.
DPG was supported by the Puzey Bequest to the University of Melbourne
and the Melbourne School of Graduate Research at the University of
Melbourne.
The work of MT is supported in part by the National Science Foundation
under grant PHY-0653563, and by NASA ATP grant NNX08AH27G.
RRV is supported in part by the Australian Research Council.


\appendix


\section{Einstein tensor components}
\label{app:einstein}

In this appendix we give explicit expressions for the non-zero
components of the Einstein tensor
\begin{equation}
G_{MN} = R_{MN} - \frac{1}{2} g_{MN} R \ ,
\end{equation}
for the case of our metric ansatz, equation~\eqref{eq:cosmo-metric}.
The indices 0 and 5 correspond to $t$ and $y$ respectively, while
both $i$ and $j$ run over the three spatial dimensions.
\begin{align}
G_{00} &= 3 \left[ \Ead \left( \Ead + \Ebd \right)
    - \frac{n^2}{b^2} \left( \Eapp + \Eap \left(\Eap - \Ebp \right) \right) \right] \ ,\\
G_{ij} &= \gamma_{ij} \frac{a^2}{b^2} \left[ \Eap \left( \Eap + 2 \Enp \right)
    - \Ebp \left( \Enp + 2 \Eap \right) + 2 \Eapp + \Enpp \right] \nonumber\\
    &\qquad + \gamma_{ij} \frac{a^2}{n^2} \left[ \Ead \left( -\Ead + 2 \End \right)
    - 2 \Eadd + \Ebd \left( -2 \Ead + \End \right) - \Ebdd \right] \ ,\\
G_{05} &= 3 \left( \Enp \Ead + \Eap \Ebd - \frac{\dot a'}{a} \right) \ ,\\
G_{55} &= 3 \left[ \Eap \left( \Eap + \Enp \right)
    - \frac{b^2}{n^2} \left( \Ead \left( \Ead - \End \right) + \Eadd \right) \right] \ .
\end{align}



\begin{thebibliography}{10}

\bibitem{ADD}
N.~Arkani-Hamed, S.~Dimopoulos, and G.~Dvali, {\it The hierarchy problem and
  new dimensions at a millimeter},  {\em Phys. Lett. B} {\bf 429} (1998) 263,
  [\href{http://xxx.lanl.gov/abs/hep-ph/9803315}{{\tt hep-ph/9803315}}].

\bibitem{Antoniadis1990}
I.~Antoniadis, {\it A possible new dimension at a few {TeV}},  {\em Phys. Lett.
  B} {\bf 246} (1990) 377--384.

\bibitem{AADD}
I.~Antoniadis, N.~Arkani-Hamed, S.~Dimopoulos, and G.~Dvali, {\it New
  dimensions at a millimeter to a fermi and superstrings at a tev},  {\em Phys.
  Lett. B} {\bf 436} (1998) 257,
  [\href{http://xxx.lanl.gov/abs/hep-ph/9804398}{{\tt hep-ph/9804398}}].

\bibitem{RS1}
L.~Randall and R.~Sundrum, {\it A large mass hierarchy from a small extra
  dimension},  {\em Phys. Rev. Lett.} {\bf 83} (1999) 3370--3373,
  [\href{http://xxx.lanl.gov/abs/hep-ph/9905221}{{\tt hep-ph/9905221}}].

\bibitem{RS2}
L.~Randall and R.~Sundrum, {\it An alternative to compactification},  {\em
  Phys. Rev. Lett.} {\bf 83} (1999) 4690--4693,
  [\href{http://xxx.lanl.gov/abs/hep-th/9906064}{{\tt hep-th/9906064}}].

\bibitem{Gibbons:1986wg}
G.~W. Gibbons and D.~L. Wiltshire, {\it {Space-Time as a Membrane in Higher
  Dimensions}},  {\em Nucl. Phys.} {\bf B287} (1987) 717,
  [\href{http://xxx.lanl.gov/abs/hep-th/0109093}{{\tt hep-th/0109093}}].

\bibitem{Visser:1985qm}
M.~Visser, {\it {An Exotic Class of Kaluza-Klein Models}},  {\em Phys. Lett.}
  {\bf B159} (1985) 22, [\href{http://xxx.lanl.gov/abs/hep-th/9910093}{{\tt
  hep-th/9910093}}].

\bibitem{Kaloper:2000jb}
N.~Kaloper, J.~March-Russell, G.~D. Starkman, and M.~Trodden, {\it {Compact
  hyperbolic extra dimensions: Branes, Kaluza-Klein modes and cosmology}},
  {\em Phys. Rev. Lett.} {\bf 85} (2000) 928--931,
  [\href{http://xxx.lanl.gov/abs/hep-ph/0002001}{{\tt hep-ph/0002001}}].

\bibitem{Starkman:2000dy}
G.~D. Starkman, D.~Stojkovic, and M.~Trodden, {\it {Large extra dimensions and
  cosmological problems}},  {\em Phys. Rev.} {\bf D63} (2001) 103511,
  [\href{http://xxx.lanl.gov/abs/hep-th/0012226}{{\tt hep-th/0012226}}].

\bibitem{Starkman:2001xu}
G.~D. Starkman, D.~Stojkovic, and M.~Trodden, {\it {Homogeneity, flatness and
  'large' extra dimensions}},  {\em Phys. Rev. Lett.} {\bf 87} (2001) 231303,
  [\href{http://xxx.lanl.gov/abs/hep-th/0106143}{{\tt hep-th/0106143}}].

\bibitem{Nasri:2002rx}
S.~Nasri, P.~J. Silva, G.~D. Starkman, and M.~Trodden, {\it {Radion
  stabilization in compact hyperbolic extra dimensions}},  {\em Phys. Rev.}
  {\bf D66} (2002) 045029, [\href{http://xxx.lanl.gov/abs/hep-th/0201063}{{\tt
  hep-th/0201063}}].

\bibitem{Rubakov:2001kp}
V.~A. Rubakov, {\it {Large and infinite extra dimensions: An introduction}},
  {\em Phys. Usp.} {\bf 44} (2001) 871--893,
  [\href{http://xxx.lanl.gov/abs/hep-ph/0104152}{{\tt hep-ph/0104152}}].

\bibitem{Langlois:2002bb}
D.~Langlois, {\it {Brane cosmology: An introduction}},  {\em Prog. Theor. Phys.
  Suppl.} {\bf 148} (2003) 181--212,
  [\href{http://xxx.lanl.gov/abs/hep-th/0209261}{{\tt hep-th/0209261}}].

\bibitem{Binetruy:1999ut}
P.~Binetruy, C.~Deffayet, and D.~Langlois, {\it {Non-conventional cosmology
  from a brane-universe}},  {\em Nucl. Phys.} {\bf B565} (2000) 269--287,
  [\href{http://xxx.lanl.gov/abs/hep-th/9905012}{{\tt hep-th/9905012}}].

\bibitem{Csaki1999}
C.~Csaki, M.~Graesser, C.~F. Kolda, and J.~Terning, {\it {Cosmology of one
  extra dimension with localized gravity}},  {\em Phys. Lett.} {\bf B462}
  (1999) 34--40, [\href{http://xxx.lanl.gov/abs/hep-ph/9906513}{{\tt
  hep-ph/9906513}}].

\bibitem{Cline1999}
J.~M. Cline, C.~Grojean, and G.~Servant, {\it {Cosmological expansion in the
  presence of extra dimensions}},  {\em Phys. Rev. Lett.} {\bf 83} (1999) 4245,
  [\href{http://xxx.lanl.gov/abs/hep-ph/9906523}{{\tt hep-ph/9906523}}].

\bibitem{Chung:1999zs}
D.~J.~H. Chung and K.~Freese, {\it {Cosmological challenges in theories with
  extra dimensions and remarks on the horizon problem}},  {\em Phys. Rev.} {\bf
  D61} (2000) 023511, [\href{http://xxx.lanl.gov/abs/hep-ph/9906542}{{\tt
  hep-ph/9906542}}].

\bibitem{Binetruy:1999hy}
P.~Binetruy, C.~Deffayet, U.~Ellwanger, and D.~Langlois, {\it {Brane
  cosmological evolution in a bulk with cosmological constant}},  {\em Phys.
  Lett.} {\bf B477} (2000) 285--291,
  [\href{http://xxx.lanl.gov/abs/hep-th/9910219}{{\tt hep-th/9910219}}].

\bibitem{Flanagan:1999cu}
E.~E. Flanagan, S.~H.~H. Tye, and I.~Wasserman, {\it {Cosmological expansion in
  the Randall-Sundrum brane world scenario}},  {\em Phys. Rev.} {\bf D62}
  (2000) 044039, [\href{http://xxx.lanl.gov/abs/hep-ph/9910498}{{\tt
  hep-ph/9910498}}].

\bibitem{Brax:2004xh}
P.~Brax, C.~van~de Bruck, and A.-C. Davis, {\it {Brane world cosmology}},  {\em
  Rept. Prog. Phys.} {\bf 67} (2004) 2183--2232,
  [\href{http://xxx.lanl.gov/abs/hep-th/0404011}{{\tt hep-th/0404011}}].

\bibitem{Takamizu:2004rq}
Y.-i. Takamizu and K.-i. Maeda, {\it {Collision of domain walls and reheating
  of the brane universe}},  {\em Phys. Rev.} {\bf D70} (2004) 123514,
  [\href{http://xxx.lanl.gov/abs/hep-th/0406235}{{\tt hep-th/0406235}}].

\bibitem{Gibbons:2006ge}
G.~Gibbons, K.-i. Maeda, and Y.-i. Takamizu, {\it {Fermions on colliding
  branes}},  {\em Phys. Lett.} {\bf B647} (2007) 1--7,
  [\href{http://xxx.lanl.gov/abs/hep-th/0610286}{{\tt hep-th/0610286}}].

\bibitem{Takamizu:2007ks}
Y.-i. Takamizu, H.~Kudoh, and K.-i. Maeda, {\it {Dynamics of colliding branes
  and black brane production}},  {\em Phys. Rev.} {\bf D75} (2007) 061304,
  [\href{http://xxx.lanl.gov/abs/gr-qc/0702138}{{\tt gr-qc/0702138}}].

\bibitem{Saffin:2007qa}
P.~M. Saffin and A.~Tranberg, {\it {The fermion spectrum in braneworld
  collisions}},  {\em JHEP} {\bf 12} (2007) 053,
  [\href{http://xxx.lanl.gov/abs/0710.3272}{{\tt arXiv:0710.3272}}].

\bibitem{Saffin:2007ja}
P.~M. Saffin and A.~Tranberg, {\it {Particle transfer in braneworld
  collisions}},  {\em JHEP} {\bf 08} (2007) 072,
  [\href{http://xxx.lanl.gov/abs/0705.3606}{{\tt arXiv:0705.3606}}].

\bibitem{Cvetic:2008gu}
M.~Cvetic and M.~Robnik, {\it {Gravity Trapping on a Finite Thickness Domain
  Wall: An Analytic Study}},  {\em Phys. Rev.} {\bf D77} (2008) 124003,
  [\href{http://xxx.lanl.gov/abs/0801.0801}{{\tt arXiv:0801.0801}}].

\bibitem{Kanti:1999sz}
P.~Kanti, I.~I. Kogan, K.~A. Olive, and M.~Pospelov, {\it {Cosmological 3-brane
  solutions}},  {\em Phys. Lett.} {\bf B468} (1999) 31--39,
  [\href{http://xxx.lanl.gov/abs/hep-ph/9909481}{{\tt hep-ph/9909481}}].

\bibitem{Mounaix2002}
P.~Mounaix and D.~Langlois, {\it {Cosmological equations for a thick brane}},
  {\em Phys. Rev.} {\bf D65} (2002) 103523,
  [\href{http://xxx.lanl.gov/abs/gr-qc/0202089}{{\tt gr-qc/0202089}}].

\bibitem{Navarro:2005uq}
I.~Navarro and J.~Santiago, {\it {Unconventional cosmology on the (thick)
  brane}},  {\em JCAP} {\bf 0603} (2006) 015,
  [\href{http://xxx.lanl.gov/abs/hep-th/0505156}{{\tt hep-th/0505156}}].

\bibitem{Rubakov:1983bb}
V.~A. Rubakov and M.~E. Shaposhnikov, {\it {Do We Live Inside a Domain Wall?}},
   {\em Phys. Lett.} {\bf B125} (1983) 136--138.

\bibitem{Akama:1982jy}
K.~Akama, {\it {An early proposal of 'brane world'}},  {\em Lect. Notes Phys.}
  {\bf 176} (1982) 267--271,
  [\href{http://xxx.lanl.gov/abs/hep-th/0001113}{{\tt hep-th/0001113}}].

\bibitem{DeWolfe:1999cp}
O.~DeWolfe, D.~Z. Freedman, S.~S. Gubser, and A.~Karch, {\it {Modeling the
  fifth dimension with scalars and gravity}},  {\em Phys. Rev.} {\bf D62}
  (2000) 046008, [\href{http://xxx.lanl.gov/abs/hep-th/9909134}{{\tt
  hep-th/9909134}}].

\bibitem{Gremm:1999pj}
M.~Gremm, {\it {Four-dimensional gravity on a thick domain wall}},  {\em Phys.
  Lett.} {\bf B478} (2000) 434--438,
  [\href{http://xxx.lanl.gov/abs/hep-th/9912060}{{\tt hep-th/9912060}}].

\bibitem{Csaki:2000fc}
C.~Csaki, J.~Erlich, T.~J. Hollowood, and Y.~Shirman, {\it {Universal aspects
  of gravity localized on thick branes}},  {\em Nucl. Phys.} {\bf B581} (2000)
  309--338, [\href{http://xxx.lanl.gov/abs/hep-th/0001033}{{\tt
  hep-th/0001033}}].

\bibitem{Kehagias:2000au}
A.~Kehagias and K.~Tamvakis, {\it {Localized gravitons, gauge bosons and chiral
  fermions in smooth spaces generated by a bounce}},  {\em Phys. Lett.} {\bf
  B504} (2001) 38--46, [\href{http://xxx.lanl.gov/abs/hep-th/0010112}{{\tt
  hep-th/0010112}}].

\bibitem{Bazeia:2007nd}
D.~Bazeia, A.~R. Gomes, and L.~Losano, {\it {Gravity localization on thick
  branes: a numerical approach}},
  \href{http://xxx.lanl.gov/abs/0708.3530}{{\tt arXiv:0708.3530}}.

\bibitem{Davies:2007xr}
R.~Davies, D.~P. George, and R.~R. Volkas, {\it {The standard model on a
  domain-wall brane}},  {\em Phys. Rev.} {\bf D77} (2008) 124038,
  [\href{http://xxx.lanl.gov/abs/0705.1584}{{\tt arXiv:0705.1584}}].

\bibitem{Davidson:2007cf}
A.~Davidson, D.~P. George, A.~Kobakhidze, R.~R. Volkas, and K.~C. Wali, {\it
  {SU(5) grand unification on a domain-wall brane from an E$_6$-invariant
  action}},  {\em Phys. Rev.} {\bf D77} (2008) 085031,
  [\href{http://xxx.lanl.gov/abs/0710.3432}{{\tt arXiv:0710.3432}}].

\bibitem{Chatillon:2006vw}
N.~Chatillon, C.~Macesanu, and M.~Trodden, {\it {Brane cosmology in an
  arbitrary number of dimensions}},  {\em Phys. Rev.} {\bf D74} (2006) 124004,
  [\href{http://xxx.lanl.gov/abs/gr-qc/0609093}{{\tt gr-qc/0609093}}].

\bibitem{Davies:2007tq}
R.~Davies and D.~P. George, {\it {Fermions, scalars and Randall-Sundrum gravity
  on domain-wall branes}},  {\em Phys. Rev.} {\bf D76} (2007) 104010,
  [\href{http://xxx.lanl.gov/abs/0705.1391}{{\tt arXiv:0705.1391}}].

\bibitem{George2007}
D.~P. George and R.~R. Volkas, {\it Kink modes and effective four dimensional
  fermion and higgs brane models},  {\em Phys. Rev.} {\bf D75} (2007) 105007,
  [\href{http://xxx.lanl.gov/abs/hep-ph/0612270}{{\tt hep-ph/0612270}}].

\bibitem{Karch2000}
A.~Karch and L.~Randall, {\it Locally localized gravity},  {\em JHEP} {\bf 05}
  (2001) 008, [\href{http://xxx.lanl.gov/abs/hep-th/0011156}{{\tt
  hep-th/0011156}}].

\bibitem{Kaloper:1999sm}
N.~Kaloper, {\it {Bent domain walls as braneworlds}},  {\em Phys. Rev.} {\bf
  D60} (1999) 123506, [\href{http://xxx.lanl.gov/abs/hep-th/9905210}{{\tt
  hep-th/9905210}}].

\bibitem{Gremm:2000dj}
M.~Gremm, {\it {Thick domain walls and singular spaces}},  {\em Phys. Rev.}
  {\bf D62} (2000) 044017, [\href{http://xxx.lanl.gov/abs/hep-th/0002040}{{\tt
  hep-th/0002040}}].

\bibitem{Afonso:2006gi}
V.~I. Afonso, D.~Bazeia, and L.~Losano, {\it {First-order formalism for bent
  brane}},  {\em Phys. Lett.} {\bf B634} (2006) 526--530,
  [\href{http://xxx.lanl.gov/abs/hep-th/0601069}{{\tt hep-th/0601069}}].

\bibitem{Flanagan:2001dy}
E.~E. Flanagan, S.~H.~H. Tye, and I.~Wasserman, {\it {Brane world models with
  bulk scalar fields}},  {\em Phys. Lett.} {\bf B522} (2001) 155--165,
  [\href{http://xxx.lanl.gov/abs/hep-th/0110070}{{\tt hep-th/0110070}}].

\bibitem{Davidson:2000hi}
A.~Davidson and P.~D. Mannheim, {\it {Dynamical localization of gravity}},
  \href{http://xxx.lanl.gov/abs/hep-th/0009064}{{\tt hep-th/0009064}}.

\bibitem{Slatyer2007}
T.~R. Slatyer and R.~R. Volkas, {\it Cosmology and fermion confinement in a
  scalar-field- generated domain wall brane in five dimensions},  {\em JHEP}
  {\bf 04} (2007) 062, [\href{http://xxx.lanl.gov/abs/hep-ph/0609003}{{\tt
  hep-ph/0609003}}].

\bibitem{Langlois:2000ia}
D.~Langlois, {\it {Brane cosmological perturbations}},  {\em Phys. Rev.} {\bf
  D62} (2000) 126012, [\href{http://xxx.lanl.gov/abs/hep-th/0005025}{{\tt
  hep-th/0005025}}].

\bibitem{Langlois:2000ns}
D.~Langlois, R.~Maartens, and D.~Wands, {\it {Gravitational waves from
  inflation on the brane}},  {\em Phys. Lett.} {\bf B489} (2000) 259--267,
  [\href{http://xxx.lanl.gov/abs/hep-th/0006007}{{\tt hep-th/0006007}}].

\bibitem{Easther2003}
R.~Easther, D.~Langlois, R.~Maartens, and D.~Wands, {\it {Evolution of
  gravitational waves in Randall-Sundrum cosmology}},  {\em JCAP} {\bf 0310}
  (2003) 014, [\href{http://xxx.lanl.gov/abs/hep-th/0308078}{{\tt
  hep-th/0308078}}].

\end{thebibliography}

\providecommand{\href}[2]{#2}\begingroup\raggedright\endgroup

\end{document}